\documentclass[submitting]{nst}

\usepackage{subfigure,dcolumn}
\usepackage[T2A,T1]{fontenc}
\usepackage[russian,english]{babel}

\usepackage{listings}
\begin{document}

\title{Detection \& imaging with Leak Microstructures}

\author{M. Lombardi}\email{Corresponding author (e-mail: {\tt mlombardi@lnl.infn.it})}
\affiliation{INFN, Laboratori Nazionali di Legnaro, viale dell'Universit\`{a} 2, Legnaro, Padova, Italy}
\author{G. Balbinot}
\affiliation{CERN, European Organization for Nuclear Research}
\author{A. Battistella}
\author{P. Colautti}
\author{V. Conte}
\author{L. De Nardo}
\affiliation{INFN, Laboratori Nazionali di Legnaro, viale dell'Universit\`{a} 2, Legnaro, Padova, Italy}
\author{G. Galeazzi}
\affiliation{Universit\`{a} degli studi di Padova, Dipartimento di Fisica Galileo Galilei, Padova, Italy}
\author{G. Prete}
\author{A. Ferretti}
\affiliation{INFN, Laboratori Nazionali di Legnaro, viale dell'Universit\`{a} 2, Legnaro, Padova, Italy}

\begin{abstract}
 Results obtained with a new very compact detector for imaging with a matrix of Leak Microstructures (LM) are reported. Spatial linearity and spatial resolution obtained by scanning as well as the detection of alpha particles with 100\% efficiency, when compared with a silicon detector, are stressed. Preliminary results recently obtained in detecting single electrons emitted by heated filament (E$_{c}$~<~1~eV) at 1-3~mbar of propane are reported.
\end{abstract}

\keywords{Gaseous detector}

\maketitle

\section{Introduction}\label{sec:1}

Some years ago we introduced a new kind of gaseous detector based on needles (or points) used as anodes, that is the Leak Microstructures (LM)~\cite{bib:1, bib:2, bib:3, bib:4, bib:5, bib:6, bib:7, bib:8}. It belongs to the family of gaseous detectors based on points as anodes such the “Detecteur Multipointes a Focalisation Cathodique”~\cite{bib:9, bib:10, bib:11}, and the “Pin Detector”~\cite{bib:12, bib:13}. The LM detector is different from the above detectors for the geometry and for some properties.

\begin{figure}[!htb]
\includegraphics
  [width=0.8\hsize]
  {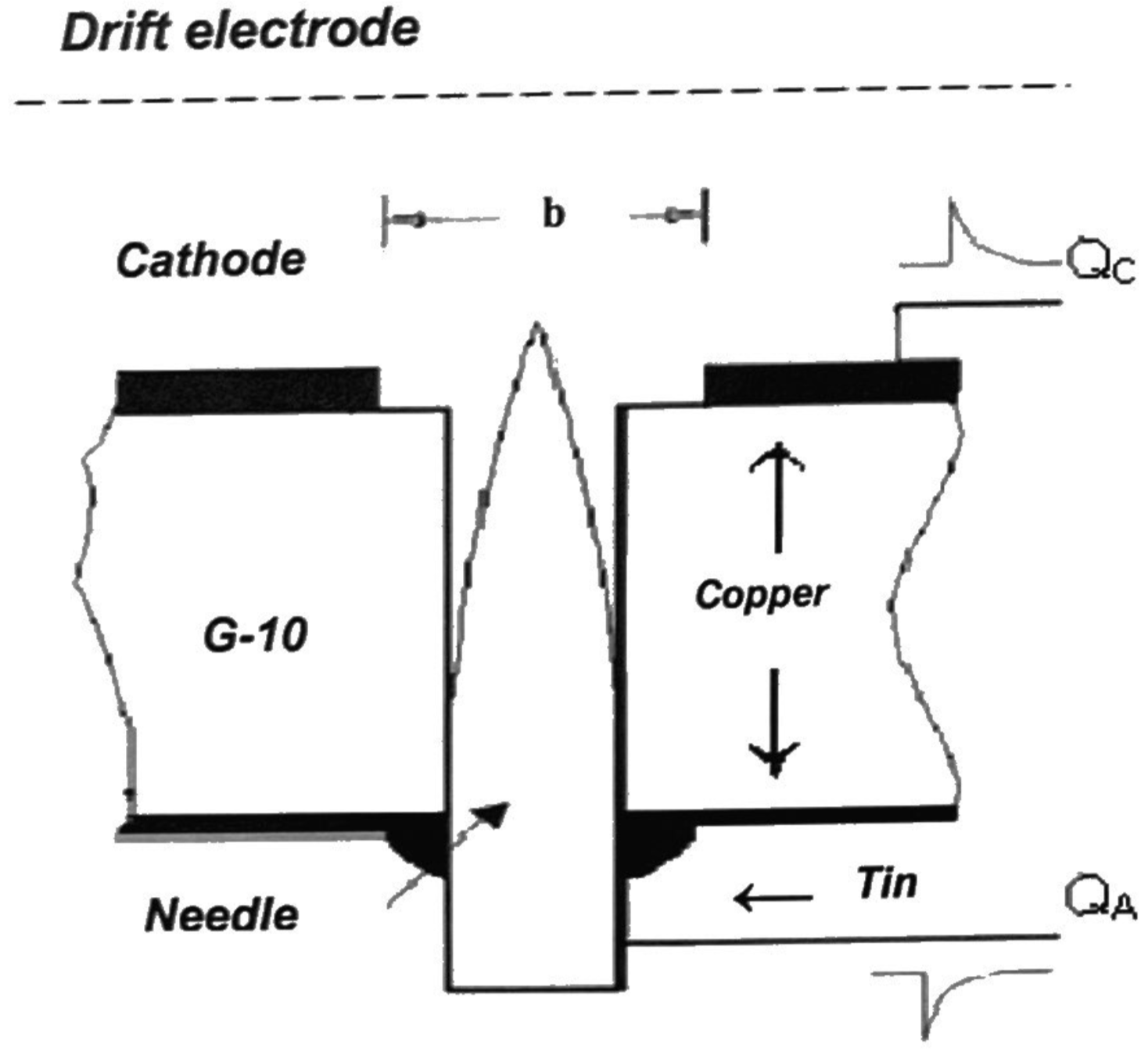}
\caption{{\it Cross section (not to scale) of a LM.}}
\label{fig:1}
\end{figure}

The geometry of a LM is really simple (fig.~\ref{fig:1}): a needle, 315~$\mu$m diameter, whose point (in the order of 20~microns) acts as anode, is inserted in a hole (0.35~mm) drilled on a vetronite supporting structure (G-10, commonly used for printed circuits), which is coppered on both sides: one side is the cathode of the detector, on the other one the needle is welded. The point of the needle emerges from the plain of the cathode 0.1-0.2~mm to work at atmospheric pressure, about 1~mm to work at low pressure~\cite{bib:14}. The copper of the cathode surrounding the needle is removed to create an insulating space (b in fig.~\ref{fig:1}) for about 550-600 $\mu$m in diameter to work at atmospheric pressure, 750-800 $\mu$m to work at low pressures (1-3 millibar). The thickness of the supporting structure, 4.5~mm, allows a very good centering of the needles to respect the insulating space of the cathode.

At distance of 3-10~mm from the cathode, a drift electrode is fixed, which marks the boundary of the active volume of the detector: this space is filled with gas. The electric field, generated between the point and the cathode, allows the primary ionization electrons multiplication. For each single ionizing radiation detected, a pair of “induced” charges (signals) are generated, one anodic and the other cathodic, with the same amplitude and time duration, of opposite sign and in time coincidence,  which are both proportional to the primary ionization. Such redundancy of information makes possible, for example, for one signal to be used to determine the energy of the incident radiation or to get timing information; the other, to provide spatial information like the point address in a matrix of LMs.

Let us shortly remember some properties of these microstructures. The absolute lack of isolating material between anode and cathode in the active volume avoids the charging-up phenomena that can alter the electric field and thus the response of the detector: the LM detector shows a very stable and repetitive behaviour.

The extreme sensitivity: it is able to detect the electrons emitted by a heated filament (E$_{c}$~<~1~eV), clearly not ionising particles~\cite{bib:7, bib:8}. The high gas gain: higher than 10$^{6}$ in detecting single electrons emitted by heated filament and more than 6$\cdot$10$^{5}$ in detecting X photons from a 5.9~keV $^{55}$Fe source~\cite{bib:6} working in proportional region.

\section{Imaging}\label{sec:2}

\begin{figure}[!htb]
\includegraphics
  [width=0.8\hsize]
  {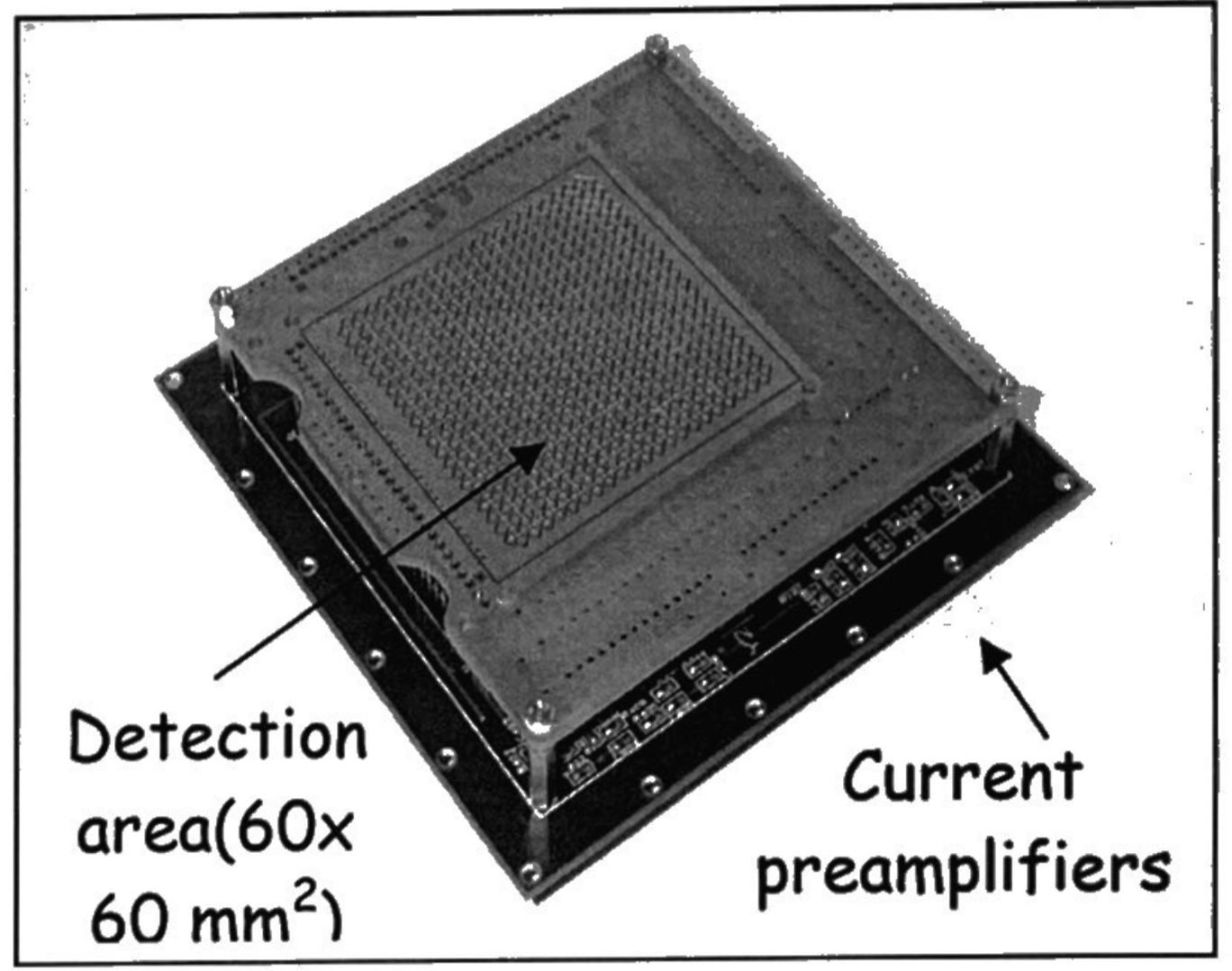}
\caption{{\it Detector with a matrix of 441 LMs and electronic set-up.}}
\label{fig:2}
\end{figure}

In fig.~\ref{fig:2} it is shown a detector with a LM matrix made by 21$\times$21 LMs (441 LMs all together), pitch 3~mm, distributed on a 60$\times$60 mm$^{2}$ surface. This last version of LMs matrix presents some innovations with regard to that described in ref.~\cite{bib:7, bib:8, bib:9, bib:10}: some experimental measures and various simulations have shown that, in order to avoid the imaging dead zones, the pitch of the needles has to be 3 mm or less.

\begin{figure}[!htb]
\includegraphics
  [width=0.8\hsize]
  {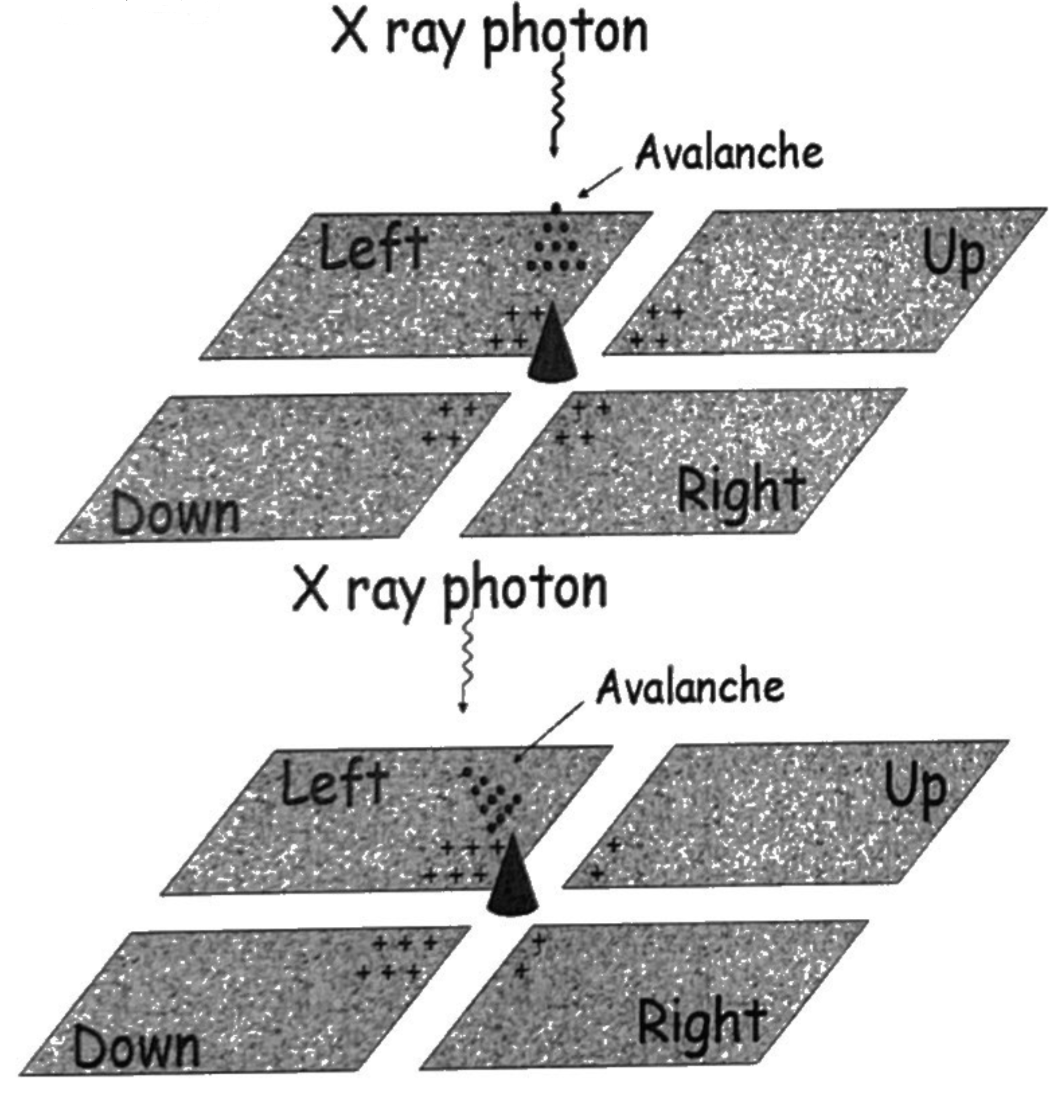}
\caption{{\it 3D scheme of a LM physical process: the cathodic charge is distributed on four pads according the position of avalanche.}}
\label{fig:3}
\end{figure}

The cathode of each LM is divided into square pads so that each microstructure is built with the point of the needle surrounded by four pads: two for the X coordinate and two for Y. This structure is particularly suitable for {\it imaging} purposes because the cathodic charge spreads on to the four pads according to the position of the striking radiation (fig.~\ref{fig:3}). To achieve a better stability of operation with the high voltage the thickness of the copper-pad- cathode, originally of 70~$\mu$m, was increased to 100~$\mu$m by galvanic deposition and has been nickel-plated in order to avoid oxidations.

\begin{figure}[!htb]
\includegraphics
  [width=0.8\hsize]
  {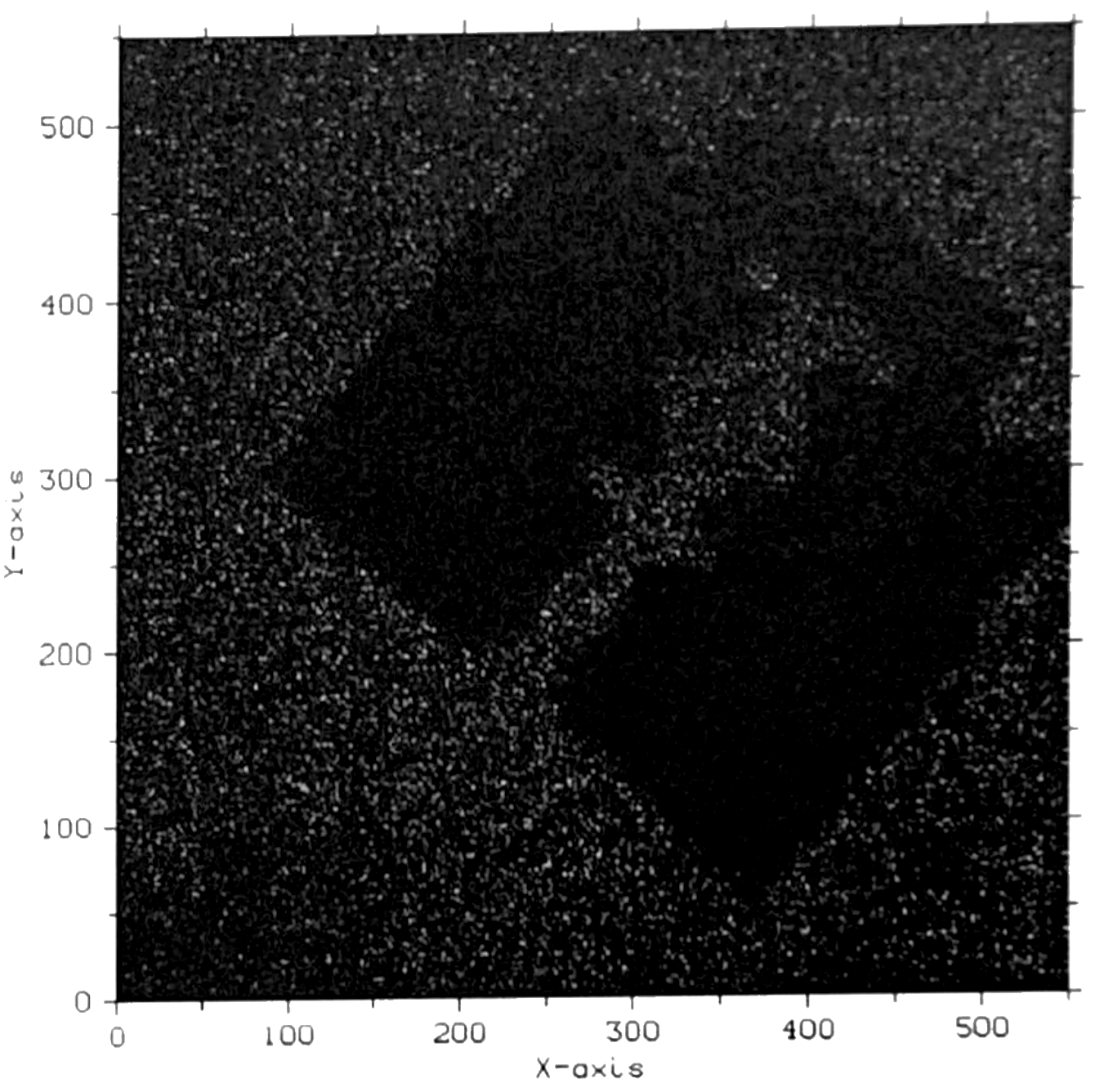}
\caption{{\it Shadowgram of a half razor-blade obtained with a matrix of LMs in 1~bar of isobutane.}}
\label{fig:4}
\end{figure}

Putting half a razor blade as mask on the drift electrode, which is fixed at a 3~mm distance from the cathode, and working in isobutane at atmospheric pressure, lightning with an extended X-ray generator, we obtain the image of fig.~\ref{fig:4}. It is important to underline that this image did not undergo any further software elaboration: it is just the presentation of the rough data. With this new detector, we can then obtain well-defined images without any imaging dead zone~\cite{bib:10}.

\begin{figure}[!htb]
\includegraphics
  [width=0.8\hsize]
  {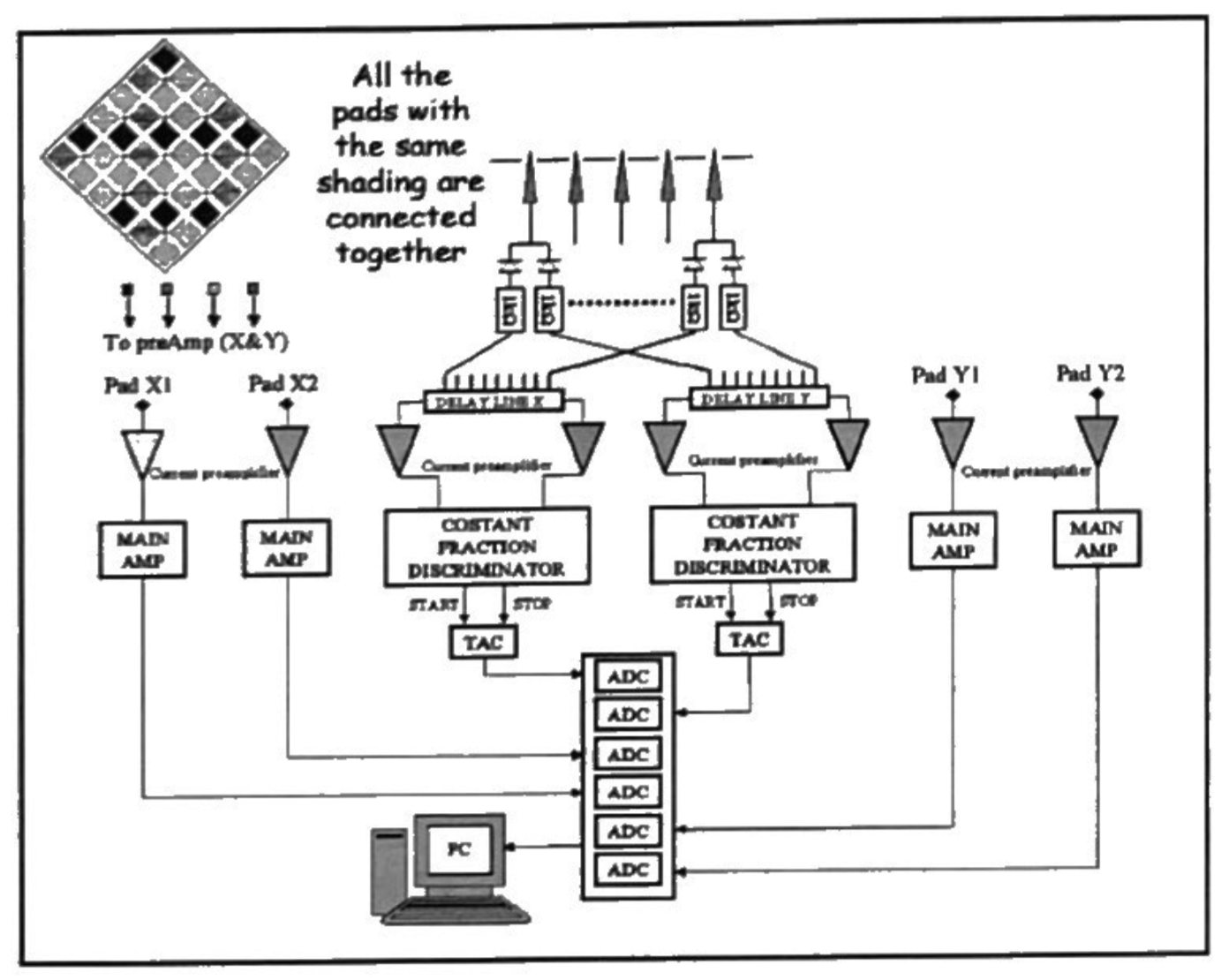}
\caption{{\it Electronic chains for imaging with a matrix of 21$\times$21 LMs.}}
\label{fig:5}
\end{figure}

The electronic chain to read-out the signals is very simple (fig.~\ref{fig:5}): as above said, each LM has four cathode pads, two for the X position (left and right) and two for the Y (up and down). We have seen that, by shortcutting all the 441 Left pads together as well as all the 441 Right, Up and Down pads (fig.~\ref{fig:5}), so that they can all be read with only four channels (4 Preamp., 4 Main Amp. and 4 ADCs), the overall spatial resolution is still very good.

The next step is to address the matrix points to get an extended image.

The charge collected by a needle (anode) is able to turn on a Shottky diode and through it to reach a preamp. Two Shottky diodes (housed in a sot23 case which is less than 3$\times$3~mm$^{2}$) are connected on the rear face of the detector to each needle in order to split the signal towards two delay lines for the addressing. With this tricky solution only six analog-to-digital converters are required: four for the pad-channels and two to read the TAC (Time to Amplitude Converter) which provides the address X,Y of the points.

\begin{figure}[!htb]
\includegraphics
  [width=0.8\hsize]
  {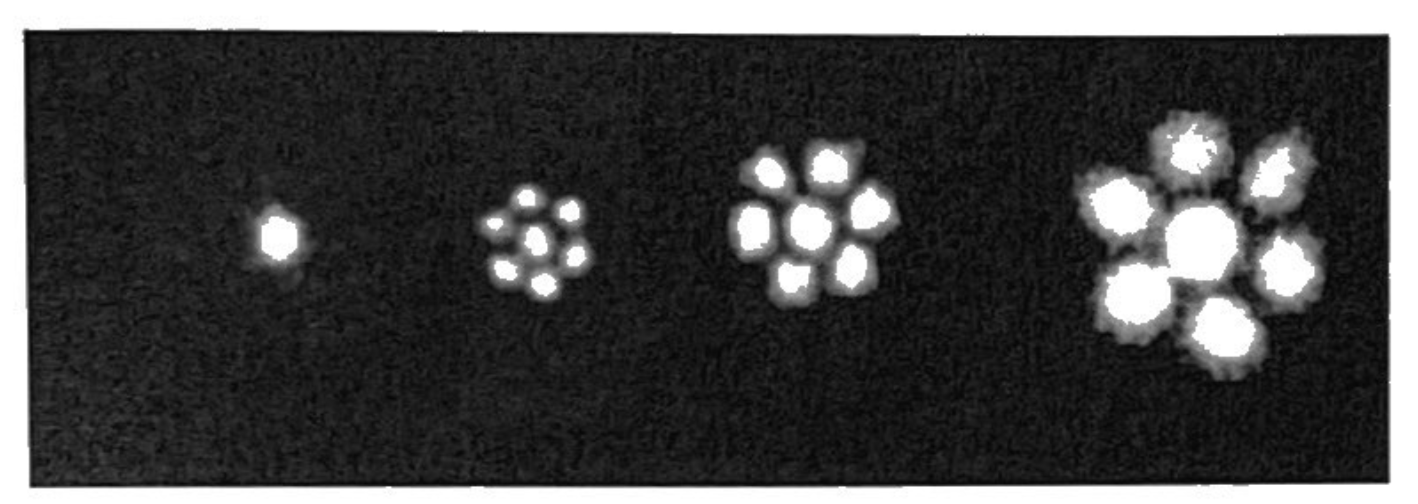}
\caption{{\it Zoom effect due to different drifting electric field; staring from left, the drifting electrode was at 50, 100, 200, 500~V, respectively.}}
\label{fig:6}
\end{figure}

The overall mean spatial resolution across the edges of fig.~\ref{fig:4}, about 460~$\mu$m FWHM, is compatible with the one of the image in fig.~\ref{fig:6}, obtained with a single LM and elsewhere described~\cite{bib:8}, taking in account that the former was obtained with a copper anti-cathode at 20 kV while the latter was obtained with a 5.9~keV $^{55}$Fe source; the mask of 7 holes, 300~$\mu$m in diameter and separated by 100~$\mu$m, was also well centred on the anode.

We can then conclude that a LM matrix detector with a 3~mm pitch allows a X-ray imaging on a 60$\times$60-mm$^{2}$ surface with a quite good spatial resolution using a very simple and compact electronic setup.

\section{Spatial linearity and spatial resolution}\label{sec:3}

\begin{figure}[!htb]
\includegraphics
  [width=0.8\hsize]
  {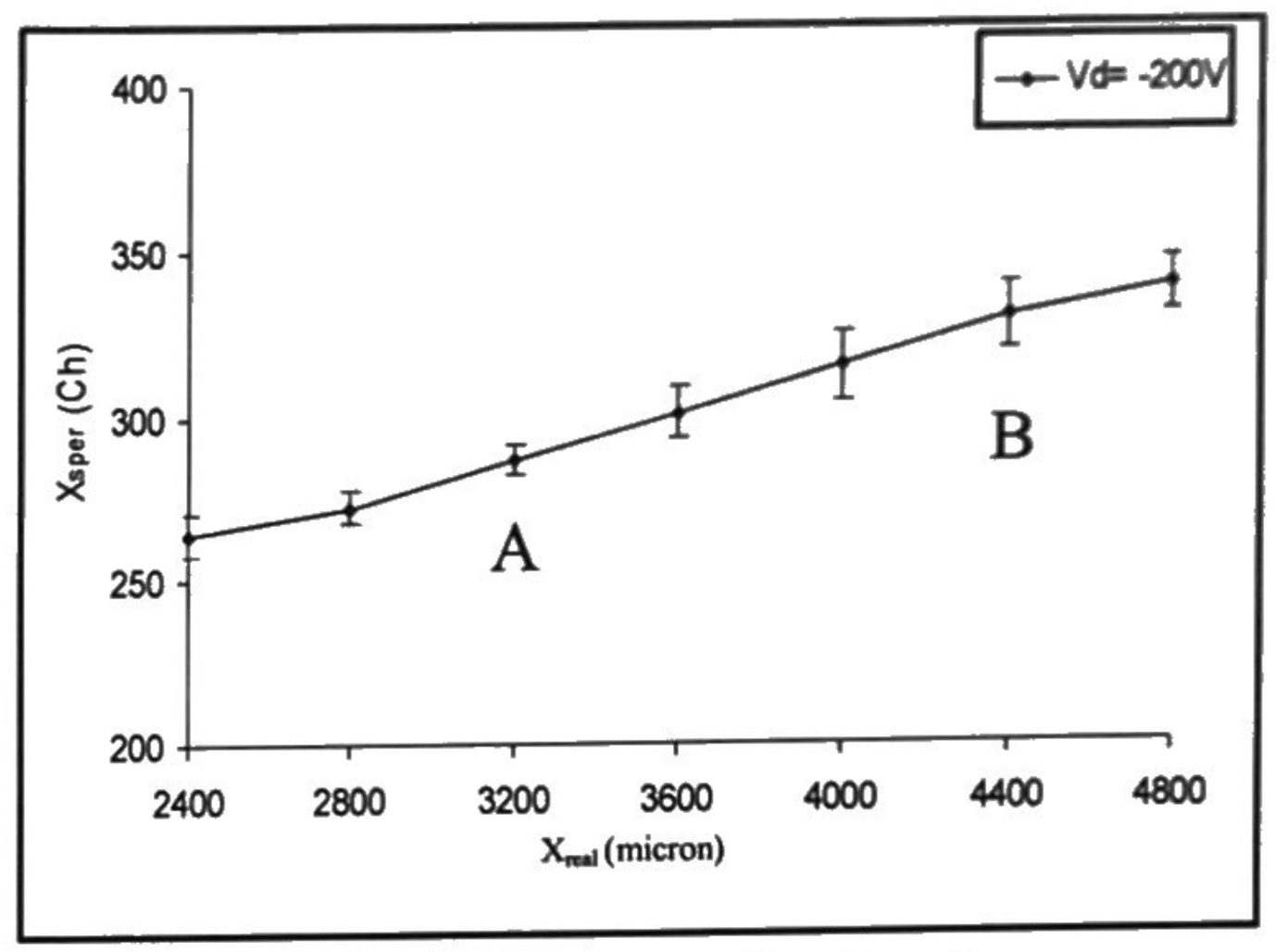}
\caption{{\it Detector position linearity.}}
\label{fig:7}
\end{figure}

Using an X-ray tube, with a copper anti-cathode at 20~kV as source, and using a LM detector with a 2.4~mm pitch we evaluated the spatial linearity in 1~bar of isobutane. A 0.1~mm slit has been put on to the drift electrode and shifted along the X-axis of constant 0.4~mm pitches. In fig.~\ref{fig:7}, we can appreciate the quite good spatial linearity: x-axis reports the micron-measured steps, whereas in the y-axis there is the position measured by the detector. The errors reported in the plot are the FWHM of the acquired data. This measure allows also the evaluation of the detector spatial resolution, as a function of the slit position respect to the anodes (points of the needles), up to the boundary between two anodes: the best spatial resolution is above an anode (A in fig.~\ref{fig:7} evaluated 279~$\mu$m FWHM) and the worse on the boundary between two anodes (B in fig.~\ref{fig:7} evaluated 643~$\mu$m FWHM).

\section{Alpha particles detection}\label{sec:4}

To evaluate the detection efficiency of alpha particles we used a single LM with an $^{241}$Am source; its efficiency is compared with a silicon detector which is known to have very high detection efficiency.

In 760~Torr of isobutane it is not possible to reach a counting plateau as a function of the increasing high voltage to the point because at this pressure alpha particles as well as X-rays emitted by the source are detected and an increase of the high voltage increases also the X rays detection efficiency.

\begin{figure}[!htb]
\includegraphics
  [width=0.8\hsize]
  {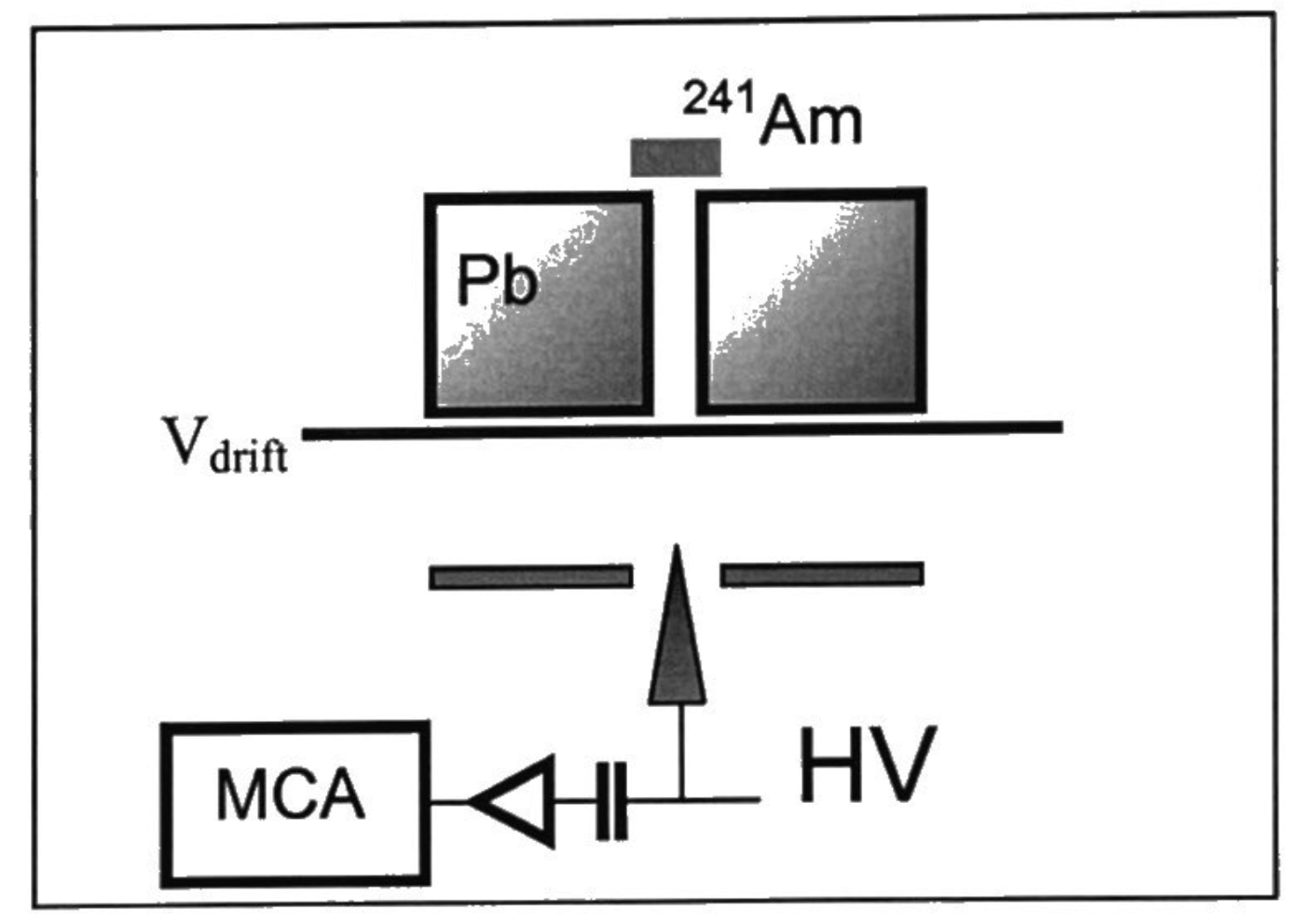}
\caption{{\it Experimental set-up.}}
\label{fig:8}
\end{figure}

At 400~Torr of isobutane, where the number of X-rays detected is negligible, we used the experimental set up of fig.~\ref{fig:8}. A lead collimator, 15~mm thick with a 3~mm diameter hole has been put at 3~mm from the cathode of a LM, well centred on the point.

\begin{figure}[!htb]
\includegraphics
  [width=0.8\hsize]
  {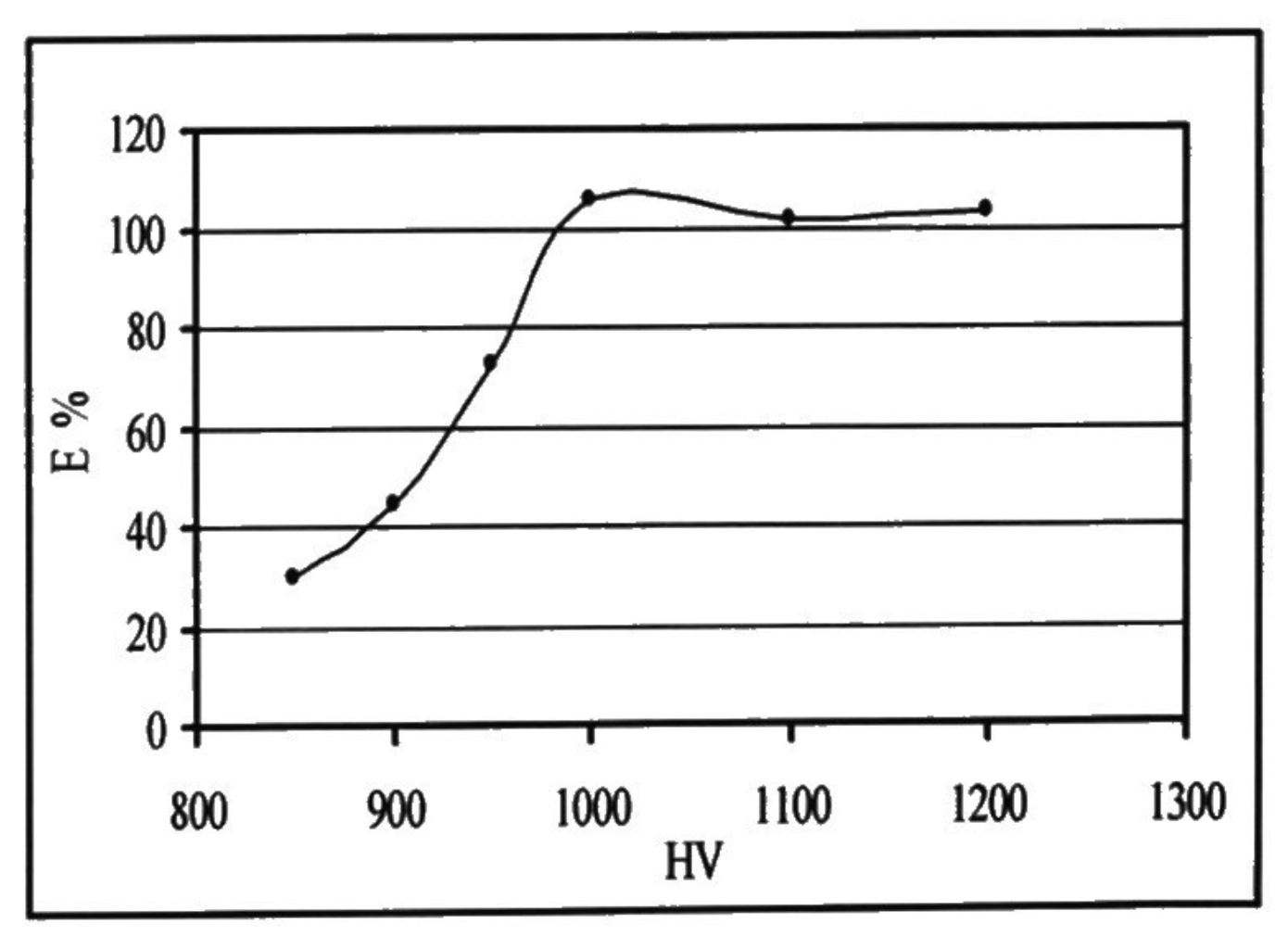}
\caption{{\it Efficiency in alpha particles detection.}}
\label{fig:9}
\end{figure}

The alpha range, under these conditions, is enough to pass the collimator and enter the active volume of the detector. In a 300~V/m drifting electric field, we obtained the curve in fig.~\ref{fig:9}: along the Y-axis is reported the relative counting efficiency to respect a silicon detector, which, when replacing the LM, is put in the output of the collimator in order to count how many alpha particles pass through it in the same lapse of measuring time. An efficiency to respect silicon detector of 100\% is reached as well as the counting plateau.

\section{Preliminary results on single electrons detection at low pressure (1-3~mbar)}\label{sec:5}

With a small matrix of 4$\times$4~LMs, pitch 2.4~mm, with the height of the point of 1~mm to respect the common-grounded-cathode, a grid acting as drift electrode 7~mm above the cathode, we detected single electrons emitted by heated filament, as well as extracted from the grid by a mercury lamp, at 1-3~mbar of propane. The insulating space surrounding the needles (b in fig.~\ref{fig:1}) was of 0.8~mm. The filament or the lamp was set above the grid.

\begin{figure}[!htb]
\includegraphics
  [width=0.9\hsize]
  {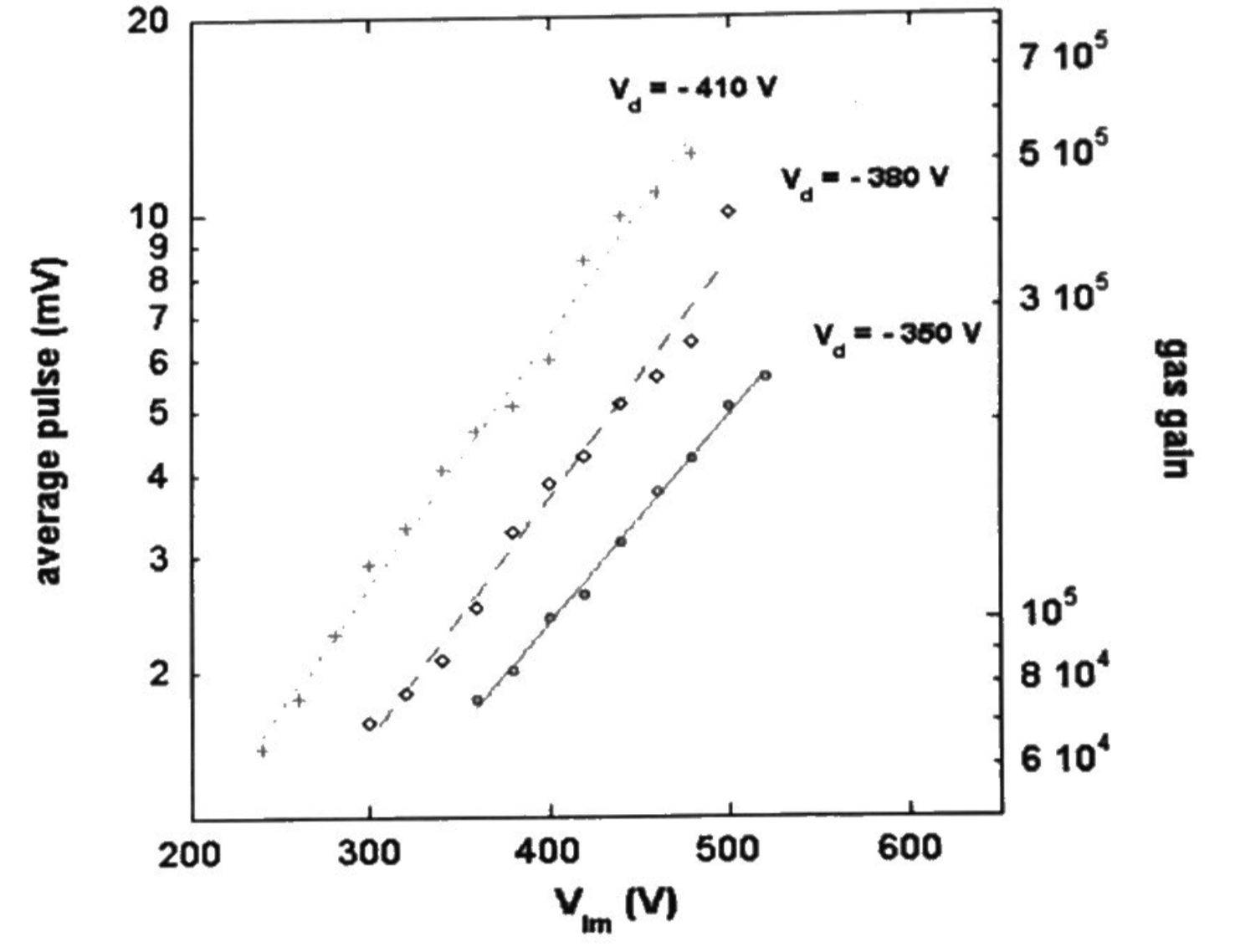}
\caption{{\it Average pulse (mV), Gas gain in 3~mbar of propane.}}
\label{fig:10}
\end{figure}

In fig.~\ref{fig:10} the average pulse in mV as well as the gas gain are reported as function of the high voltage applied to the common points (V$_{lm}$) for three different voltages applied to the grid (V$_{d}$) at 3~mbar of propane. As calculated in ref.~\cite{bib:14} at 3~mbar of propane, with the height of the point of 1~mm, the Townsend ionisation coefficient a is bigger than 1 everywhere in the active volume of the detector starting from the grid: the avalanche spreading depends on the voltage applied to the grid (V$_{d}$), which therefore will influence not only the drift but also the process of multiplication.

The current preamp. used has 15~mV/$\mu$A dynamic characteristic. Signals were well recorded also at 1~mbar of propane. The behaviour in proportional region is evident.

\begin{figure}[b]
\includegraphics
  [width=0.9\hsize]
  {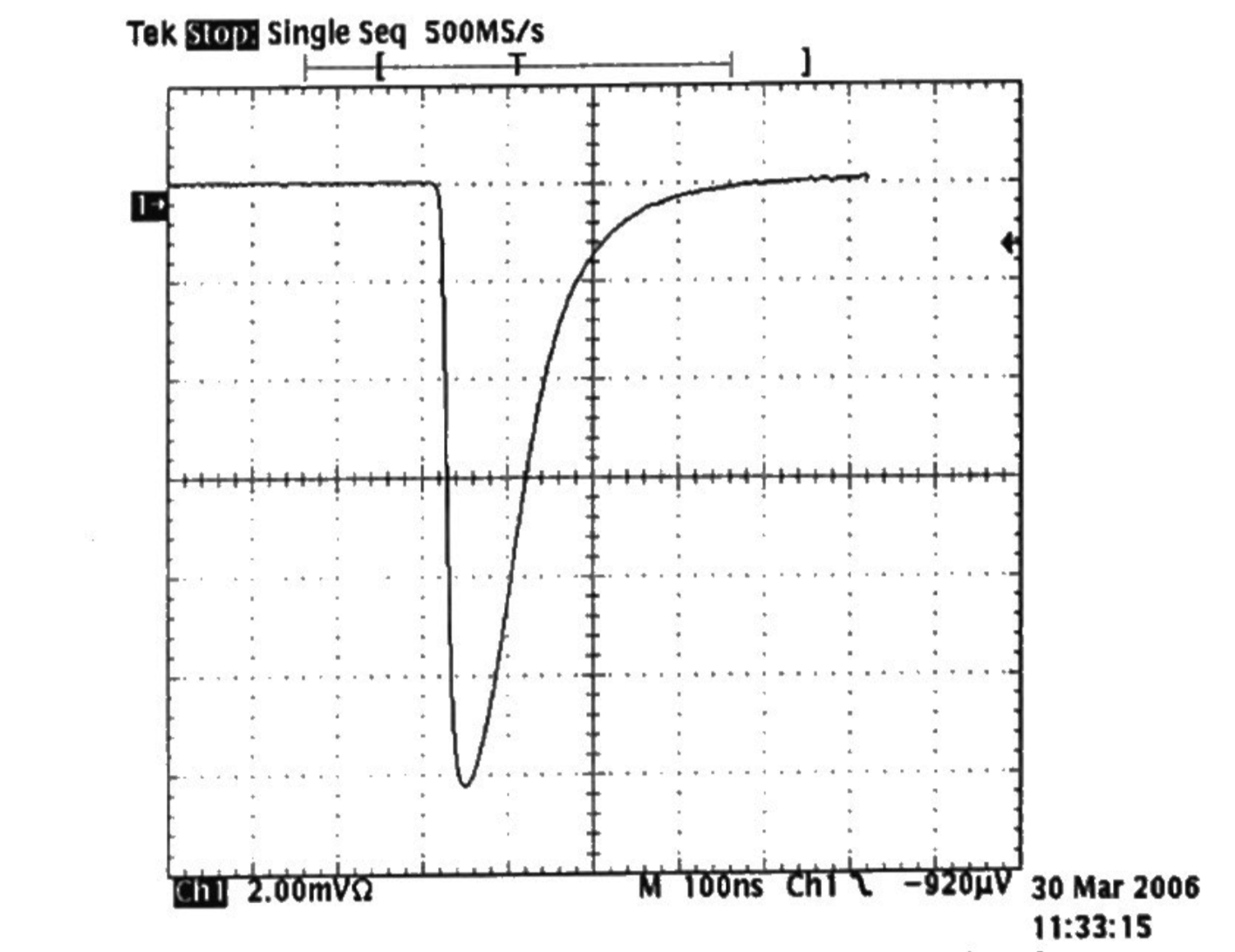}
\caption{{\it Average of pulses of single electron detected in 3~mbar of propane.}}
\label{fig:11}
\end{figure}

In fig.~\ref{fig:11} an example of the average of pulses recorded by a LeCroy 8300A digital oscilloscope at V$_{lm}$~480~V;V$_{d}$~410~V in 3~mbar of propane. In order to test if the leak microstructure (LM) counter can be used to detect single-electron in STARTRACK nanodosimeter, single-electron pulse-height spectra were measured in propane gas at low pressure (3~mbar)~\cite{bib:18}. Experimental data show good prospects for this single step detector: LM detects single-electrons operating in proportional mode also at low pressure, the pulse-height spectra are well fitted by theoretical Polya distribution, allowing to calculate the single-electron multiplication efficiency, which can reach a value of 96\%, quite similar to that of the MSAC detector proposed by A. Breskin.
\\

\section{Conclusions}\label{sec:6}

Up to now we developed the part of multiplication of primary electrons and the rough handling of the signals obtained in a very simple compact and reliable way to obtain images with quite good spatial resolution but with low X-ray conversion efficiency. In fact the conversion efficiency of soft X rays in 1~bar of isobutane is less than 1\%~\cite{bib:15}. Next part of this job, to enhance the detection efficiency, is to use a suitable photocathode which could be of Secondary Electrons Emission (SEE) type~\cite{bib:16, bib:17} or some other type. The detection capability of single electrons at very low gas pressure (1-3~mbar) working in proportional region is achieved.

\end{document}